# The Atmospheric Extinction of Light

*Stephen W. Hughes[1], Michael Cowley[2,3]*
*Sean Powell[1], Joshua Carroll[1]*

*[1]Department of Chemistry, Physics and Mechanical Engineering,*
*Queensland University of Technology, Gardens Point Campus,*
*Brisbane, Queensland 4001, Australia*
*[2]Department of Physics & Astronomy,*
*Macquarie University, Sydney, NSW 2109, Australia*
*[3]Australian Astronomical Observatory,*
*PO Box 915, North Ryde, NSW 1670, Australia*

**Abstract**
An experiment is described that enables students to understand the properties of atmospheric extinction due to Rayleigh scattering. The experiment requires the use of red, green and blue lasers attached to a traveling microscope or similar device. The laser beams are passed through an artificial atmosphere, made from milky water, at varying depths, before impinging on either a light meter or a photodiode integral to a Picotech Dr. DAQ ADC. A plot of measured spectral intensity verses depth reveals the contribution Rayleigh scattering has to the extinction coefficient. For the experiment with the light meter, the extinction coefficient for red, green and blue light in the milky sample of water were 0.27, 0.36 and 0.47 $cm^{-1}$ respectively and 0.032, 0.037 and 0.092 $cm^{-1}$ for the Picotech Dr. DAQ ADC.

## Introduction

In astronomy, information is collected from a variety of wavelengths – i.e. cosmic rays, x-rays, ultraviolet, visible, infrared, radio. Only certain wavelengths can pass through the atmosphere with little attenuation. For example, air molecules in the atmosphere cause sunlight to scatter in all directions. During the day, the shorter wavelength blue component of sunlight is scattered in our atmosphere creating the blue colour of the sky. However, at dusk or dawn, the sunlight must traverse more atmosphere, scattering more blue light and leaving us with red-orange sunsets (Strutt 1871).

The main mechanism is through elastic Rayleigh scattering, which is dependent on the inverse fourth power of wavelength. Therefore, light with a wavelength of 400 nm is scattered over nine times more than 700 nm photons. Rayleigh scattering also occurs in space, as seen in figure 1, which is a photo of the Pleiades star cluster taken by Austrian astronomer, Johannes Schedler. In this case, the Pleiades stars are illuminating foreground dust (Gibson and Nordsieck 2003). The blue appearance results from blue light being scattering more effectively by small dust particles.

When light passes through the atmosphere, some is absorbed and some scattered. Since a constant fraction of the photons in the primary beam are removed per unit thickness of material, the intensity of the beam falls exponentially according to Beer's law:

$$I(\lambda) = I_0(\lambda) e^{-kx}$$

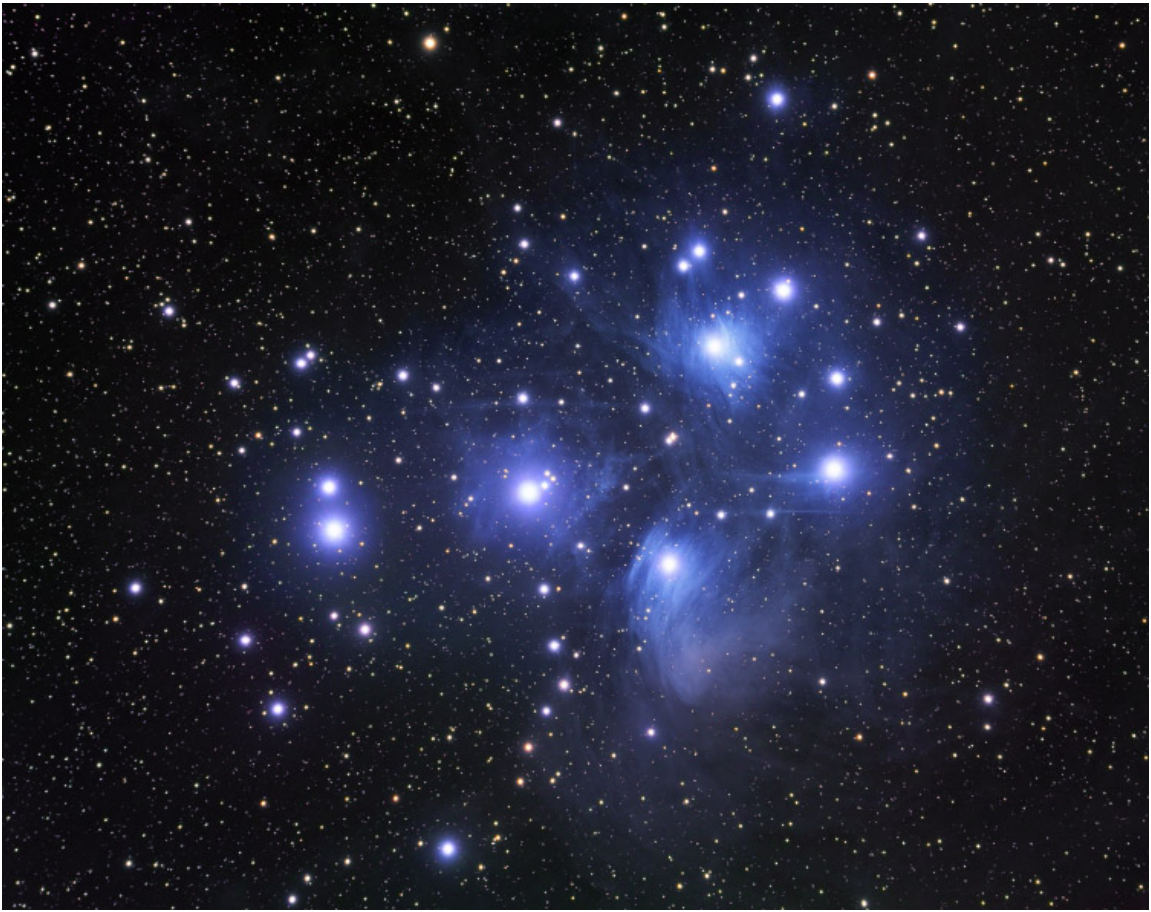

Figure 1. Blue light scattered by dust in front of the Pleiades star cluster. Image courtesy of Johannes Schedler (www.panther-observatory.com)

$I_0(\lambda)$ is the initial spectral intensity (radiant intensity per unit wavelength), $I(\lambda)$ is the spectral intensity after the beam has traversed the material, $k$ is the extinction coefficient (including absorption and scattering) and $x$ is the thickness of material

Beer's law only applies to measurements with monochromatic radiation, such as lasers. The cost of lasers, especially blue lasers, has fallen in the last few years and they can now be obtained for 10's of US dollars.

In this experiment the extinction (or absorption and scattering) of red, green and blue light by milky water is measured. Milky water represents an artificial atmosphere, where milk proteins act as an analogue to air molecules. The inspiration behind this experiment is a paper by Mahmood bin Mat (1992). However, there is no shortage of similar experiments that have tested the effects of extinction by small particles (e.g. Goulden 1958, Sakurada and Nakamura 2002, Gedzelman *et al* 2008; Gedzelman and Vollmer 2011)

**Method**
The wavelengths of the lasers used are red 635 nm, green 532 nm and blue 473 nm. The power of the red and green lasers was 1 mW and the blue laser 5 mW. Figure 2 shows the basic experimental setup.

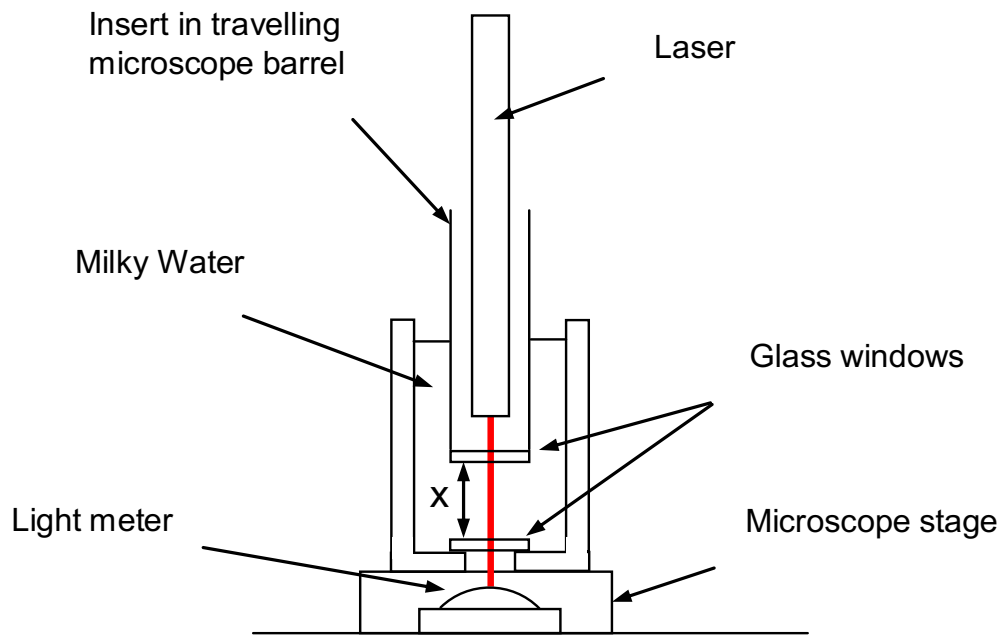

Figure 2. Schematic diagram of the apparatus. The depth of water ($x$) is varied by raising and lowering the barrel.

A Perspex tank is filled with water and a small drop of milk added. Some trial and error might be required so that the milky water is transparent enough for the blue laser to penetrate. Lasers are placed in a holder in the barrel of the traveling microscope (with the optics removed). A plastic ring and piece of paper is used to depress the laser switch to keep it in the on position. For the first experiment described in this paper, an Emtek EMT-201 light meter was placed directly beneath the glass window of the tank. This is a photometric light meter and so measures intensity as perceived by the human eye. The barrel of the traveling microscope was lowered so that it touched the window at the bottom of the Perspex tank.

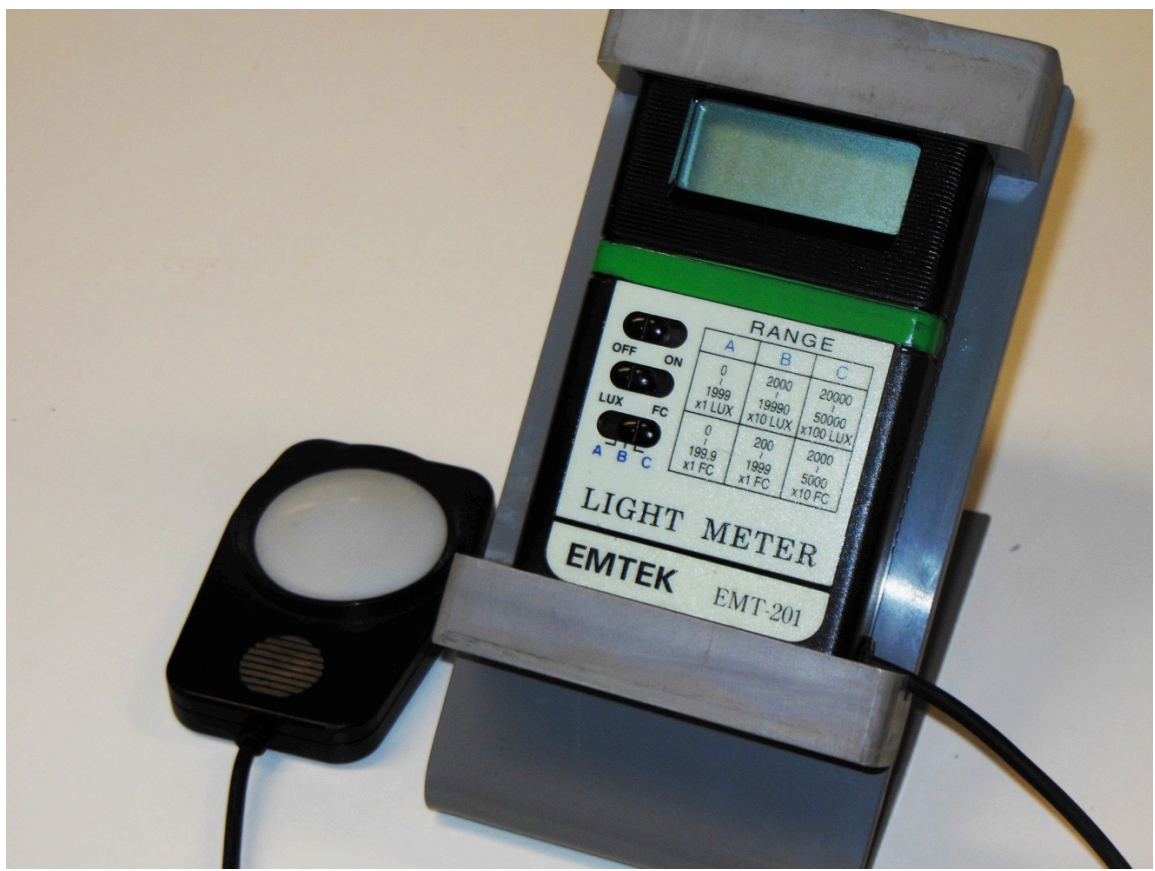

Figure 3. Emtek light meter used to obtain the data for the first part of the experiment shown in table 1 and plotted in figure 5.

A wire was used to remove any bubbles stuck to the window of the microscope barrel containing the laser. The focussing knob of the travelling microscope was adjusted so that the barrel was raised in 5 mm increments. At each level the spectral intensity was recorded. The barrel was raised until it was just below the surface of the water. The experiments were performed in a dark room to reduce interference from ambient light.

The experiment was repeated for a different milky water sample with intensity measured using a Picotech Dr. DAQ ADC (figure 4) with an inbuilt photodiode, which in essence is a radiometric light meter. The Dr. DAQ is a combined analogue to digital converter, spectrum analyser and function generator that is connected to a PC via a USB connection. The ADC is 8-bit with a bandwidth of 100 kHz.

The photodiode chip is only a few mm across and comparable in size to the width of the laser beam, making it very difficult to keep the laser aligned with the chip as the barrel of the microscope is raised. To reduce the effect of the variation in the position of the laser, a piece of tracing paper was placed between the tank and Dr.DAQ to serve as a diffuser.

The PicoScope software was set to record the photodiode signal over a period of several minutes (please see the video abstract). The barrel was raised by 5 mm and maintained at that position for about 10 s. At the end of each run the data was stored in comma separated values (CSV) format so it could be read into spreadsheet software such as Microsoft Excel. The waveforms for each laser were loaded into the spreadsheet software and the cursor moved to the middle of each step to obtain an intensity reading.

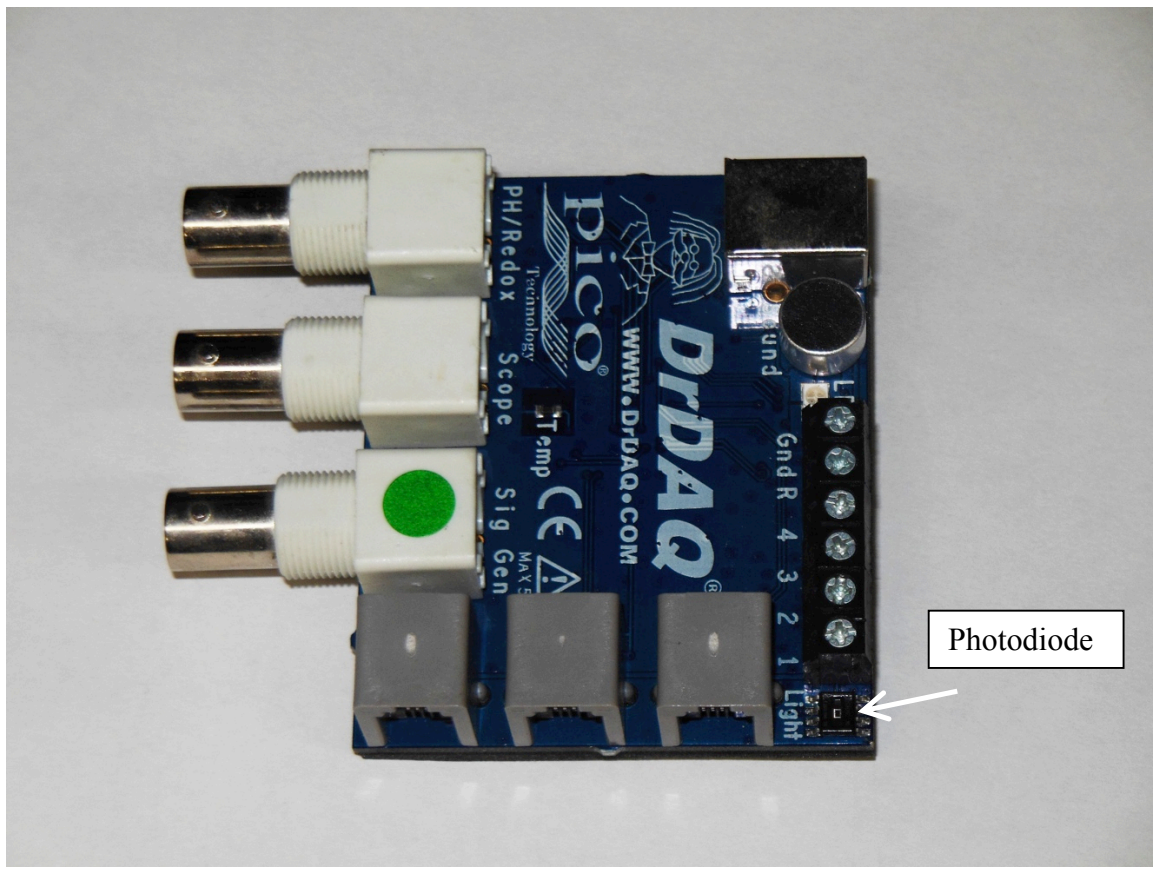

Figure 4. Light metre on a Dr. DAQ ADC used to collect data for the second part of the experiment shown in table 2 and plotted in figures 6 and 7.

## Results

Table 1 shows the data obtained using an Emtek light meter. Figure 5(a) is a plot of spectral intensity verses depth. Intensities were normalised by dividing intensity by the intensity for zero depth ($I/I_o$). Figure 5(b) shows a plot of the natural logarithm of the normalised intensity verses depth.

The attenuation coefficient for each colour is the gradient of the line. For the milky water used in the Emtek experiment the respective attenuation coefficients for red, green and blue were 0.27, 0.36 and 0.47 cm$^{-1}$. The correlation coefficients of the straight lines fitted to the natural logarithm plots were 0.9996, 0.9999 and 0.9974 for red, green and blue respectively. These can be taken as a measure of the how closely Beer's law is followed.

Table 2 shows the data obtained using the Dr. DAQ photodiode. Figure 6 shows a plot of the basic data obtained with the Dr. DAQ photodiode, figure 7(a) is a plot of normalised spectral intensity versus depth for the Dr.DAQ photodiode, and figure 7(b) a plot of the natural logarithm of the normalised intensity data. The respective attenuation coefficients for red, green and blue were 0.032, 0.037 and 0.092 cm$^{-1}$. These values are not expected to be the same as those obtained for the Emtek as different milky water was used. The correlation coefficients for the lines fitted to the natural log plots were 0.9957, 0.9774 and 0.9317. In this case there is a marked deviation from Beer's law for blue.

Table 1. Intensity values taken with an Emtek meter. The error in each reading is ± 0.5 units.

| depth (cm) | red | green | blue |
|---:|---:|---:|---:|
| 0 | 500 | 3120 | 51 |
| 0.5 | 440 | 2630 | 40 |
| 1 | 389 | 2200 | 30 |
| 1.5 | 341 | 1830 | 24 |
| 2 | 297 | 1529 | 20 |
| 2.5 | 259 | 1275 | 15 |
| 3 | 226 | 1059 | 12 |
| 3.5 | 198 | 882 | 9 |
| 4 | 172 | 737 | 7 |
| 4.5 | 150 | 609 | 6 |
| 5 | 129 | 506 | 5 |

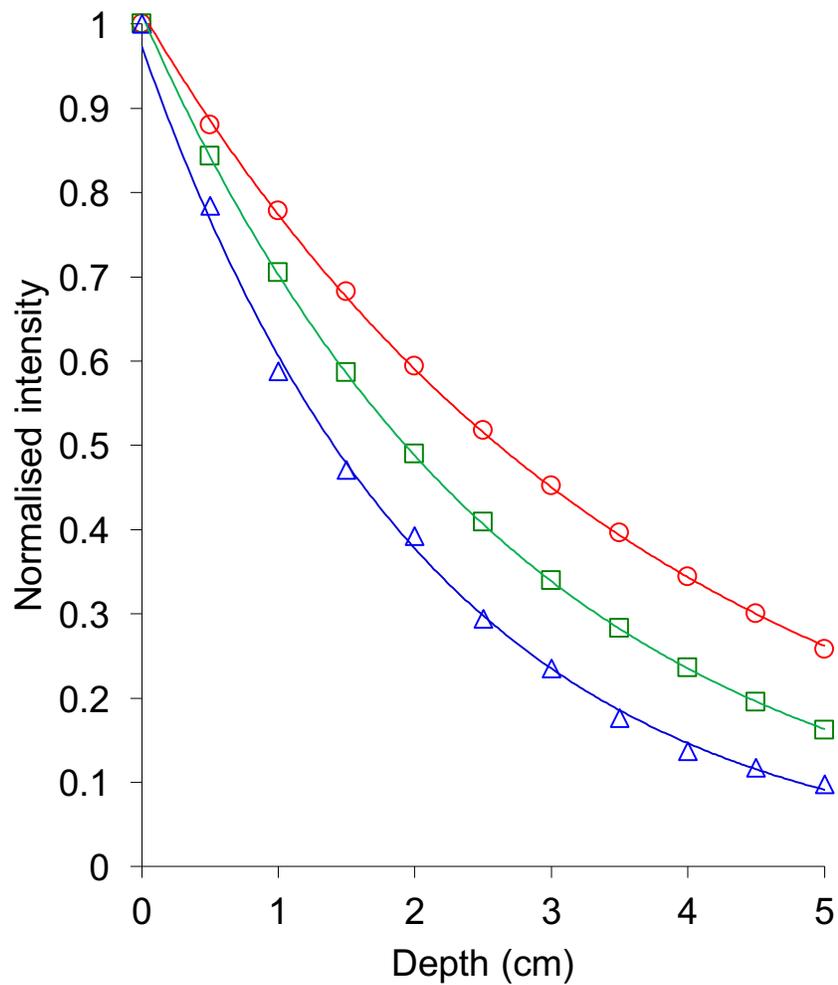

Figure 5. (a) Plot of normalised laser intensity values versus depth taken with an Emtek meter. In this plot and all subsequent plots, circles correspond to the red laser, squares the green and triangles the blue. Error bars have not been drawn as they vary between 0.1% of the reading for the zero depth and 10% for the lowest blue reading, and therefore range in size from being completely invisible to a maximum height about the same as the height of the data markers. (b) Plot of the natural logarithm of normalised intensity versus depth.

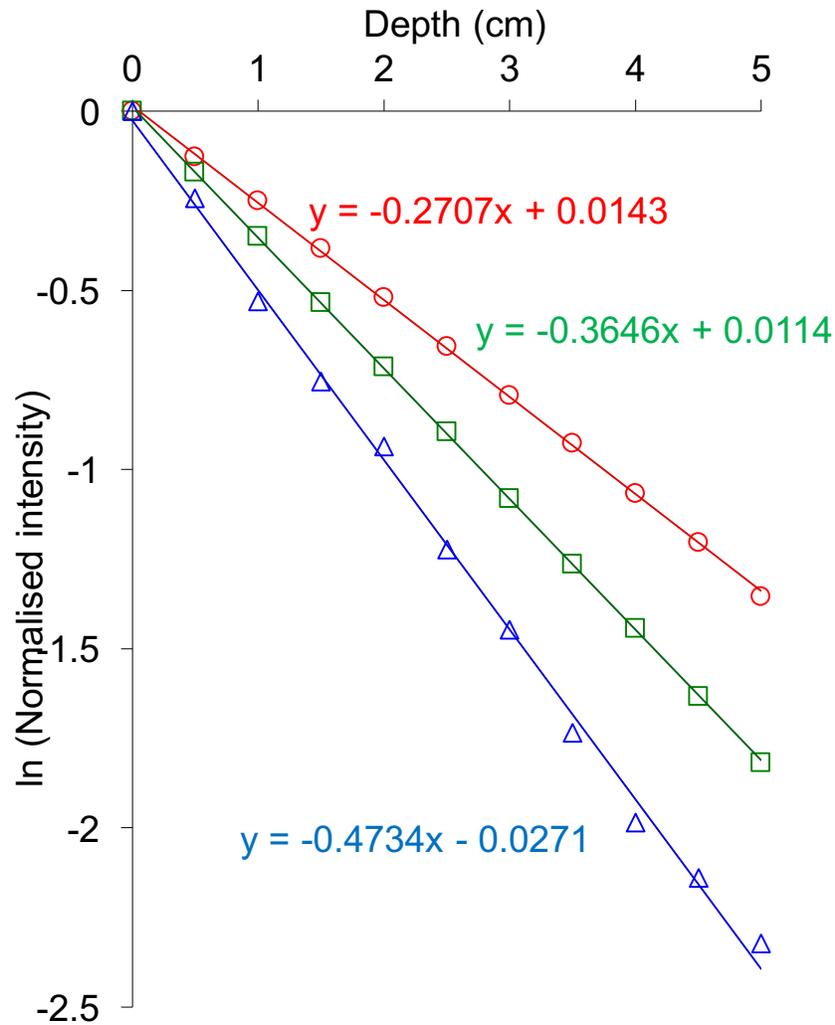

Figure 5. (b) Plot of the natural logarithm of normalised laser intensity versus depth.

To assess the noise amplitude of the Dr. DAQ photodiode data, the mean and standard deviation was calculated for a sequence of 10 measurements taken from the middle of the third step from the origin in the data (figure 6). The standard deviations, expressed as a percentage of the mean are 0.85%, 0.71% and 2.14% for the red, green and blue lasers respectively. If error bars were added to figures 7(a) and (b) they would be smaller than the markers.

Table 2. Laser intensities taken using the Dr. DAQ ADC photodiode. The data were obtained by plotting the data in Excel and then placing the cursor in the middle of each step to extract the value of the closest point.

| depth (cm) | red | green | blue |
|---|---|---|---|
| 0 | 68.1 | 50.3 | 21.7 |
| 0.5 | 59.4 | 34.2 | 9.1 |
| 1 | 49.1 | 28.3 | 5.2 |
| 1.5 | 40.5 | 22.9 | 3.2 |
| 2 | 34.3 | 19.4 | 2.4 |
| 2.5 | 29.7 | 16.9 | 2.1 |
| 3 | 25.6 | 13.8 | 1.9 |
| 3.5 | 22.7 | 12.1 | |
| 4 | 19.6 | 10.7 | |

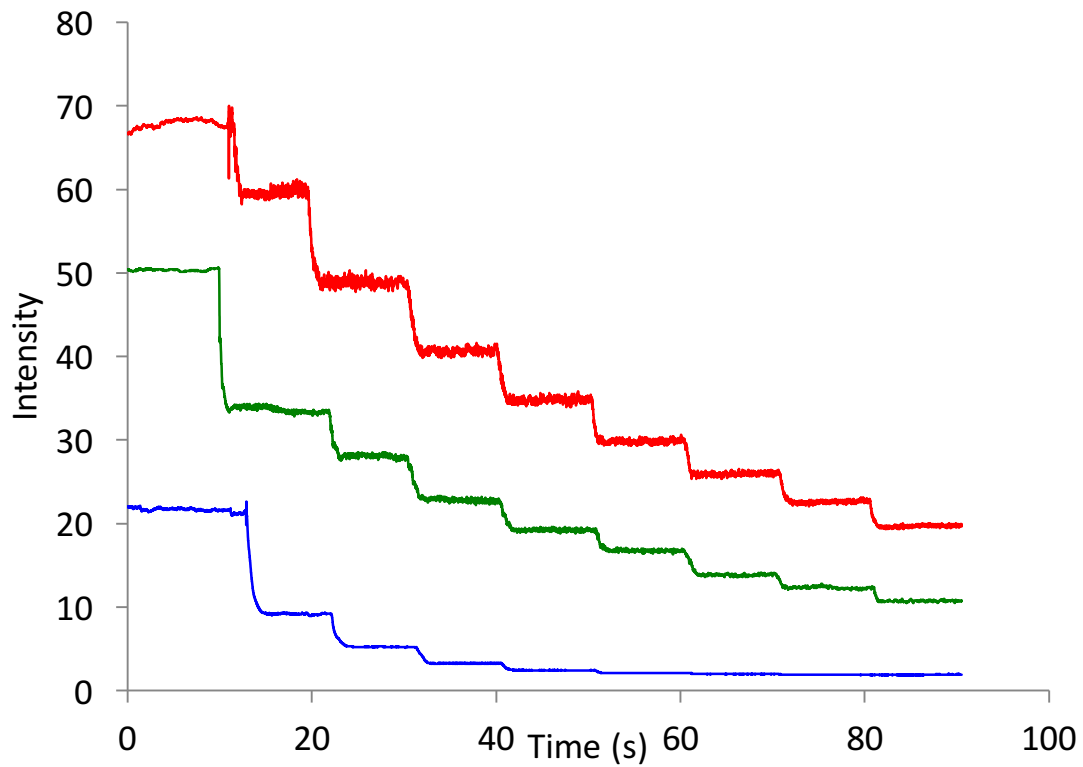

Figure 6. Plot of laser intensities verses time taken with a Dr. DAQ ADC with an inbuilt photodiode. The steps are due to an increase in the depth of water of 0.5 cm.

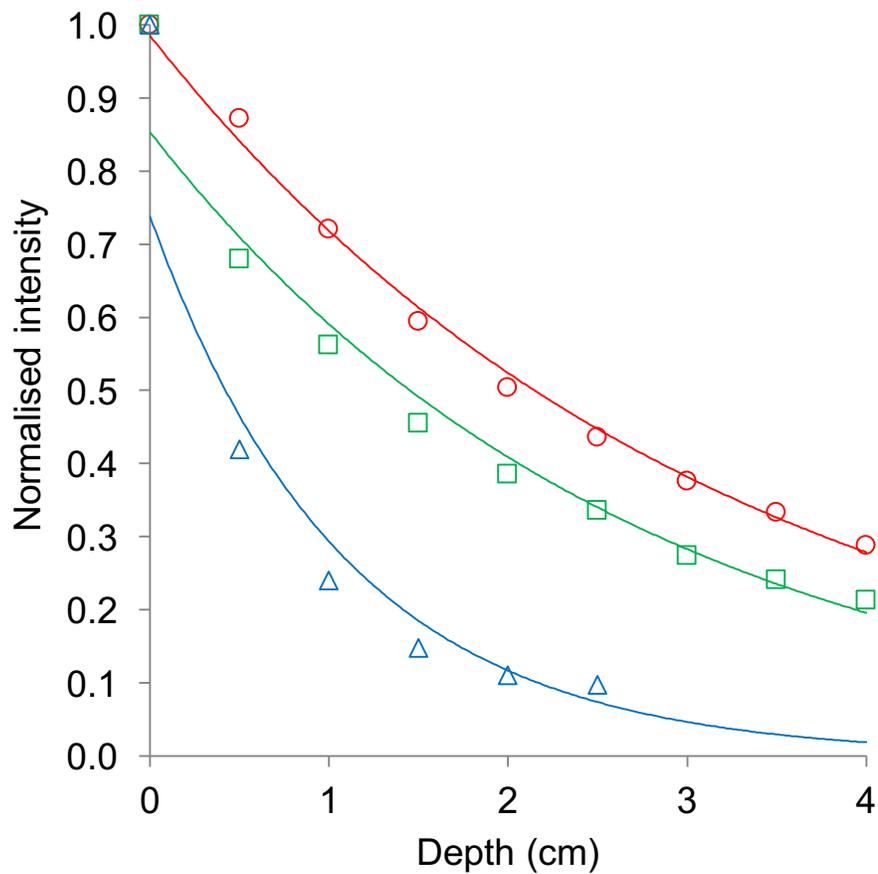

Figure 7. (a) Plot of normalised laser intensity values versus depth taken with a Dr. DAQ ADC photodiode. If error bars were added as they would be smaller than the markers. (b) Plot of laser intensities versus time taken with a Dr. DAQ ADC inbuilt photodiode. The steps are due to an increase in the depth of water of 0.5 cm

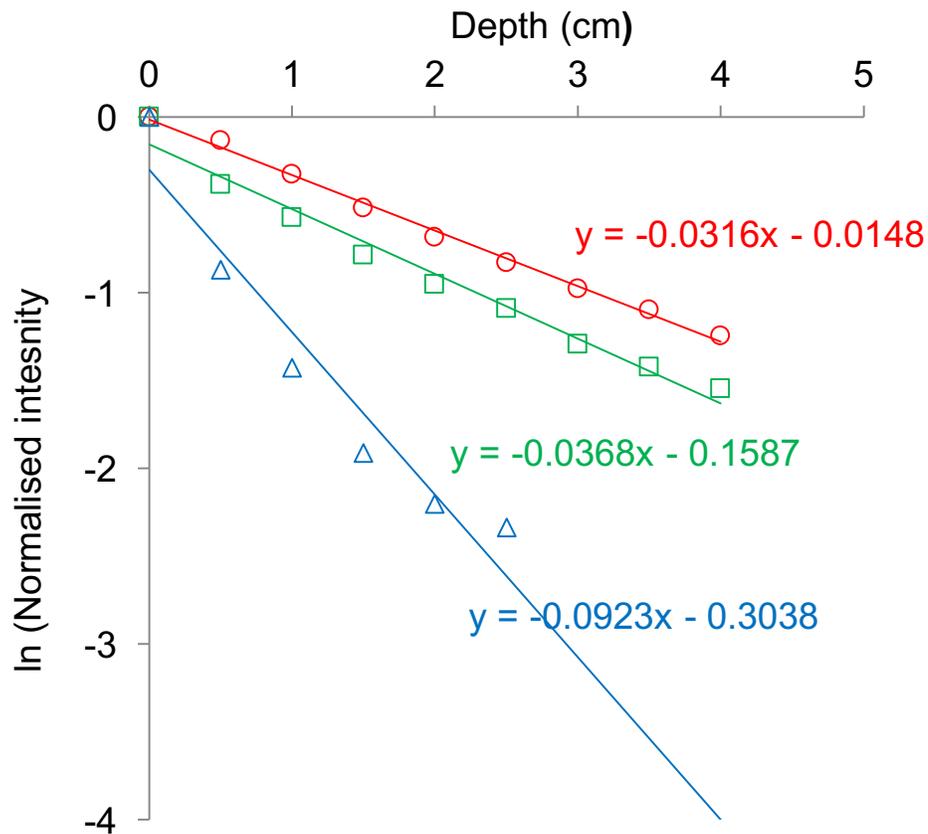

Figure 7. (b) Plot of laser intensities verses time taken with a Dr.DAQ ADC inbuilt photodiode. The steps are due to an increase in the depth of water of 0.5 cm.

**Discussion**

The activity described in this paper represents a basic approach to demonstrate the properties of atmospheric extinction due to Rayleigh scattering. By using milky water as an analogue for Earth's atmosphere, we show the extinction coefficient for blue light is higher than for green, which in turn was higher than for red. This extinction is dominated by Rayleigh scattering from the small milk proteins in the milky water, and explains the observation of the blue sky during the day and the red-orange sky at dusk or dawn.

We demonstrate the experiment and produce results from two light measurement devices, an Emtek light meter and Dr. DAQ photodiode. Even although one meter is photometric and the other radiometric, reasonable data were obtained. However, the Emtek data were superior to the Dr. DAQ photodiode data. The reason is likely due to significant scattering of the green and blue light in the diffuser, which degrades the pristine Beer's law response seen with the Emtek data. The quality of the photodiode data could be improved by having a small aperture at the bottom of the tank to reduce the amount of scattered light falling on either the Emtek or Dr. DAQ. Although not related to astrophysics, this apparatus lends itself for studying the properties of turbid media (Gao *et al* 2013). The educational advantage of using the Dr. DAQ ADC is that students can see on the computer screen that blue light is attenuated more than green and red light.

Under certain circumstances, Beer's law breaks down due to limitations associated with chemical and instrumental factors. The most prominent of these limitations are:

(i) Stray ambient light on the detector. Such light has the potential to degrade the quality of the data, but can typically be mitigated by performing the experiment in a darkened room.

(ii) Change in concentration over time. Milk proteins in the dilute solution may clump over time, reducing the optical thickness and varying the extinction coefficient. To mitigate this effect, measurements should be taken as promptly as possible.

(iii) Excessive concentration of milk in solution. If the concentration of milk in the dilute solution is too high, it increases the optical thickness. This can lead to multiple scattering events, where Beer's law breaks down. As detailed in the method, trail and error is required to achieve a suitable concentration.

Safety is an important issue (Astronomical Society of Australia, Fact Sheet number 22). Lasers with a power of 1 mW or less are not considered dangerous. However, many lasers labelled 1 mW have a much higher power. For example, author SWH was given a '1 mW' green laser that turned out to be 50 mW. So, it would be advisable to check the power of the lasers intended for use in an extinction experiment. In our case the red and green lasers were class II (< 1mW) and the blue laser class IIIa (<5 mW). The red and green laser were regular laser pointers but the blue laser was more powerful and therefore the students were required to do an online safety course and answers questions that were marked by the university laser safety officer. Students were required to pass the test before being allowed to do the experiment.